# A 5.9 GHz Sezawa SAW Acoustic Delay Line Based on $Al_{0.6}Sc_{0.4}N$-on-Sapphire with Propagation $Q$-factor > 3,000

Chin-Yu Chang, Xiaolei Tong, Pedram Yousefian, Ella Klein, Xingyu Du, and *Roy H. Olsson III
Department of Electrical and Systems Engineering, University of Pennsylvania, PA, USA
*rolsson@seas.upenn.edu

*Abstract*—In this work, we demonstrate a high-performance surface acoustic wave (SAW) delay line based on a Scandium alloyed aluminum nitride (AlScN)-on-sapphire platform operating at 5.9 GHz with an exceptionally high acoustic propagation $Q$-factor. An 800 nm AlScN thin film with 40% scandium alloying concentration was deposited on a thick sapphire substrate to achieve strong acoustic energy confinement and large electromechanical coupling effect, thereby minimizing the insertion loss (IL) and propagation loss (PL) of the acoustic delay line (ADL). The proposed ADL was designed to operate in the Sezawa mode using a Single-Phase Unidirectional Transducer (SPUDT) electrode configuration for better unidirectionality. The fabricated ADLs with different delay lengths, after conjugate matching, exhibited delay times spanning 13 to 214 ns and IL ranging from 7.6 to 18.3 dB. The extracted PL reached as low as 9.2 dB/mm at 5.9 GHz, with a group velocity ($v_g$) of around 5,779 m/s. Based on these results, the proposed ADLs exhibit a high acoustic propagation $Q$-factor of 3,044. These findings highlight the potential of AlScN-on-sapphire platforms for high operational frequency, low-loss SAW ADL devices in advanced RF applications.

## I. INTRODUCTION

The rapid evolution of next-generation wireless communications system has created a growing demand for compact, low power consumption, and low-loss delay line components. True-time-delay (TTD) networks are essential for phased-array beamforming in massive multiple-input multiple-output (MIMO) communication systems, where they enhance spectral efficiency, suppress co-channel interference in dense deployments, and support low-latency services [1]. In parallel, full-duplex radio system require high performance radio frequency components. Besides circulators that isolate the transmitter and receiver path, cancelling circuits also rely on reconfigurable and low-loss delay synthesis elements to emulate multipath channels and suppress strong self-interference [2], [3], [4]. Together, these applications highlight the critical need for delay line technologies that combine low insertion loss (IL), wide bandwidth, and scalability toward multi-GHz frequencies to enable future wireless systems with higher capacity, improved energy efficiency, and enhanced robustness against interference.

Several approaches have been explored to realize high-frequency delay elements, including electromagnetic (EM) waveguides and active delay circuits based on complementary metal-oxide-semiconductor (CMOS) technologies. EM-based delay lines can provide fine delay resolution and reconfigurability, but they suffer from large physical size due to the long wavelength, which significantly lowers the achievable delay density. CMOS-based circuits can generate tunable delays and even provide signal gain, yet they inherently consume power, typically offer only picosecond-scale delays, and require complex circuit topologies that occupy substantial chip area [5], [6], [7]. On the other hand, acoustic delay lines (ADLs) exploit the much shorter wavelength of elastic waves to realize passive devices with inherently high delay density, which makes them compact, low loss, and scalable toward higher frequency band operation. These unique attributes enable ADLs to support applications such as low-noise oscillators [8], [9], [10], [11], sequentially switched non-reciprocal networks [12], [13], integrated phononic circuits [14], [15], phase modulators [16] and acoustoelectric amplifiers [17], [18].

ADLs have long been implemented using surface acoustic wave (SAW) technology with bulk piezoelectric materials. In this approach, metallic interdigital transducers (IDTs) are patterned on the piezoelectric substrate to launch traveling acoustic waves, and the IDT pitch directly defines the wavelength and operational frequency through lithographic processes [19]. However, conventional SAW ADLs on bulk piezoelectric substrates suffer from substantial acoustic energy leakage, which results in high propagation loss (PL) and large insertion loss (IL) as the frequency increases, ultimately limiting their scalability for high-frequency operation [20]. With the advancement of thin-film piezoelectric technologies, suspended membrane waveguides supporting plate waves, i.e., Lamb wave, were developed to reduce the acoustic energy leakage and minimized both PL and IL while retaining the intrinsic advantages of acoustic propagation [21], [22]. Yet, these devices suffer from complex fabrication processes, limited mechanical robustness, and are highly sensitive to fabrication variance. In contrast, thin-film SAW ADLs utilize the acoustic velocity mismatch between the piezoelectric layer and the substrate to minimize energy leakage and enhance electromechanical coupling efficiency. These features have propelled the performance of SAW resonant transducers toward next-generation high-frequency applications [23], [24].

Thin-film SAW devices have been implemented with various piezoelectric materials on high-stiffness or waveguided substrates. Typical combinations include lithium niobate (LN), lithium tantalate (LT), and scandium-alloyed aluminum nitride (AlScN) integrated with sapphire [25], silicon carbide [26], [27], [28], quartz [29] , diamond [30], and piezoelectric-on-insulator (POI) [31], [32], [33]. LN and LT provide very high electromechanical coupling and are suitable for wide-bandwidth applications, but their integration still relies on thin-film transfer processes that are complex and difficult to scale for ultrathin layers. AlScN, on the other hand, provides a boost in the electromechanical coupling factor compared to aluminum nitride (AlN) and is compatible with direct deposition by physical vapor deposition (PVD), enabling high-quality thin films with thicknesses down to the tens-of-nanometer range [34], [35]. Moreover, for ADL applications the c-axis oriented AlScN film is essentially isotropic in the propagation plane. This eliminates the concern of power flow angle (PFA) encountered in anisotropic crystals such as LN and LT. This intrinsic isotropy provides greater flexibility for device layout and simplifies the design of delay lines.

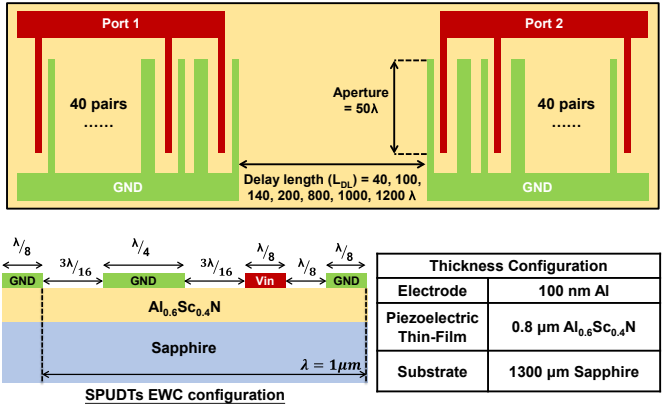

Fig. 1. Schematic and the detail design parameters of the proposed AlScN-on-Sapphire SAW delay line at 5.9 GHz.

Despite considerable progress on ADLs operating above 5 GHz, most studies have focused on plate wave-based ADLs [36], [37], [38], [39], [40], [41], [42], [43]. In comparison, studies on high-frequency thin-film SAW ADLs employing AlScN remain very limited. In this work, we demonstrate a thin-film SAW ADL on an AlScN-on-sapphire platform. The device excites the Sezawa mode at 5.9 GHz with single-phase unidirectional transducers (SPUDTs), as shown in Fig. 1. Various different delay lengths ($L_{DL}$) are fabricated to characterize the ADL performance of this platform. The proposed ADL attains exceptionally low PL and IL across a wide delay range, thereby achieving a remarkably high acoustic propagation $Q$-factor over 3,000. These results establish AlScN-on-sapphire as a promising platform for next-generation high-frequency delay lines and RF applications.

## II. ALSCN-ON-SAPPHIRE ACOUSTIC DELAY LINE

The proposed ADL device was realized on the AlScN-on-sapphire platform and designed to operate at its Sezawa mode. Sapphire was selected as the substrate material in this platform owing to its high stiffness and affordability. A high-stiffness substrate introduces an acoustic velocity mismatch between the piezoelectric layer and the substrate, which improves acoustic energy confinement. For the piezoelectric thin-film, $Al_{0.6}Sc_{0.4}N$ was chosen for its higher piezoelectric coefficient compared to AlN. The Sezawa mode is known for its high electromechanical coupling factor compared to the Rayleigh mode in AlScN-based thin-film SAW devices [27], [30]. Moreover, the Sezawa mode can offer higher phase velocity which is beneficial for frequency scaling.

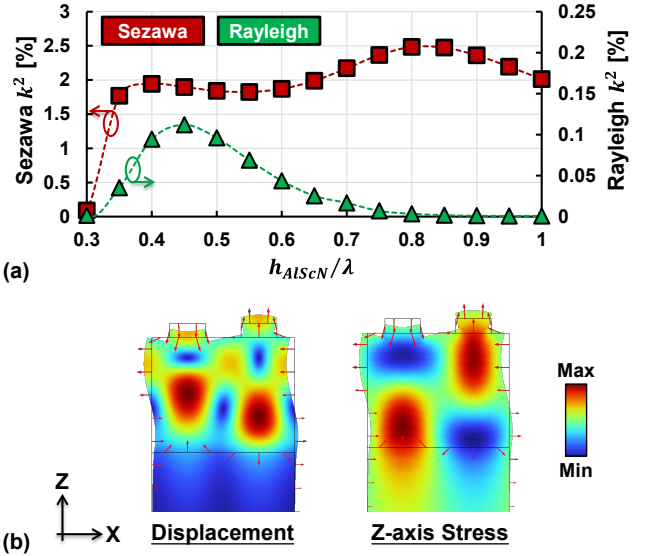

Fig. 2. (a) Simulated intrinsic electromechanical coupling factor ($k^2$) of Rayleigh and Sezawa mode with different AlScN thickness-to-wavelength ($h_{AlScN}/\lambda$) ratio. (b) Simulated displacement and stress profile in z-axis of the Sezawa mode with Al electrode.

To optimize the performance of the proposed ADL, high electromechanical coupling factor is essential, as it enables a broader fractional bandwidth (FBW) while simultaneously reducing the IL. The enhanced coupling factor strengthens the reflectivity of the transducer electrodes, which mitigates bidirectional propagation and further relaxes the intrinsic trade-off between FBW and IL of the ADLs [44], [45]. Therefore, a 2D periodic unit cell model is implemented in finite element analysis (FEA) to optimize the piezoelectric thin-film thickness-to-wavelength ratio ($h_{AlScN}/\lambda$). The intrinsic coupling coefficient ($k^2$) was extracted from the simulated phase velocities under electrically open ($v_{p\_open}$) and electrically short ($v_{p\_short}$) boundary conditions for both Rayleigh and Sezawa modes, as expressed in (1).

$$k^2 = \frac{v_{p\_open}^2 - v_{p\_short}^2}{v_{p\_open}^2} \qquad (1)$$

The simulation results indicate that the highest $k^2$ occurs at $h_{AlScN}/\lambda = 0.8$ while the $k^2$ of Rayleigh mode becomes very weak, as shown in Fig. 2(a). A similar trend has also been reported in [46]. This design regime not only secures the performance of the Sezawa mode but also facilitates the spectrally clean ADLs. Note that the material properties of the $Al_{0.6}Sc_{0.4}N$ used in the simulation are obtained from [47]. Fig. 2(b) shows the simulated displacement and z-axis stress profile of the Sezawa mode from the unit cell model with an 800 nm of $Al_{0.6}Sc_{0.4}N$ on a sapphire substrate and a 100 nm thick Al electrode.

In addition to optimizing $k^2$, the electrode design is also critical to the performance of ADLs. A conventional interdigital transducer (IDT), with quarter-wavelength line and width spacing, launches acoustic waves in both directions and thus suffers from an intrinsic 6-dB bi-directional loss. To overcome this drawback, single-phase unidirectional transducers (SPUDTs) have been developed. By incorporating reflector electrodes within each unit cell, SPUDTs enhance forward propagation and suppress backward propagation. This constructive and destructive interference mechanism effectively minimizes IL. Among various SPUDT configurations, the electrode-width-controlled (EWC) design is particularly suitable for wideband ADLs. The EWC-SPUDT offers relatively weaker reflectivity but provides a broader FBW. This feature helps mitigate the intrinsic trade-off between FBW and insertion loss, enabling the ADL to achieve wideband operation with acceptable transmission efficiency. After defining the electrode configuration, the proposed ADLs were designed with an aperture of 50λ and 40 pairs for both port transducers. To characterize the intrinsic PL of the Sezawa mode in the AlScN-on-sapphire platform, devices with different delay lengths ranging from 40λ to 1,200λ were fabricated, allowing IL to be evaluated as a function of propagation distance.

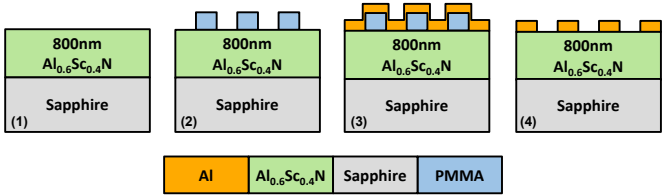

Fig. 3. Fabrication process flow in this work.

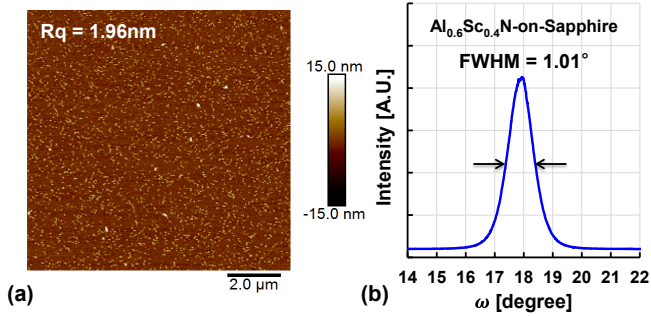

Fig. 4. (a) AFM results of the 800nm $Al_{0.6}Sc_{0.4}N$ thin-film deposited on the sapphire substrate showing a root mean square surface roughness (Rq) of 1.96nm. (b) Rocking curve of the deposited AlScN thin-film measured by X-ray diffraction (XRD). The measured Full Width at Half Maximum (FWHM) is 1.01°.

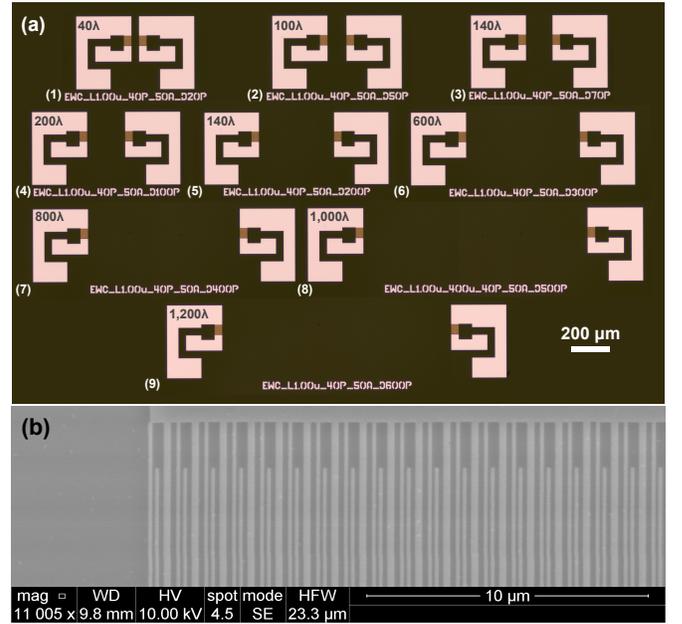

Fig. 5. (a) Stitched microscopic image of fabricated ADL devices with different delay length. (b) SEM image of the fabricated electrode with SPUDT configuration.

Fig. 3 illustrates the fabrication process flow of the proposed ADLs. (1) The process started with a 6-inch single-side polished c-plane sapphire wafer. An $Al_{0.6}Sc_{0.4}N$ thin film was deposited by sputtering using an Evatec CLUSTERLINE® 200 II physical vapor deposition (PVD) system. The deposition included an AlN seed layer, an AlScN gradient layer, and a bulk $Al_{0.6}Sc_{0.4}N$ layer [27], [48]. The deposited film was first characterized using atomic force microscopy (AFM), yielding a root-mean-square surface roughness (Rq) of 1.96 nm over a $10 \times 10$ μm$^2$ area. The crystal quality was examined by X-ray diffraction (XRD). A rocking curve of ~1° confirmed the highly c-axis oriented growth. (2) The deposited wafer was diced into pieces with a dicing saw to prepare for lithography. Electron-beam (e-beam) lithography (Raith EBPG5200) was used to define the ADL patterns on a single-layer PMMA A4 950 resist. (3) After development, a 100-nm Al layer was deposited using e-beam evaporation. (4) Finally, the e-beam resist was stripped out by photoresist remover to finish the lift-off process.

Fig. 5(a) shows the stitched microscopic image of the fabricated ADLs with delay lengths from 40λ to 1200λ. Fig. 5(b) presents a scanning electron microscope (SEM) image confirming the successful definition of the SPUDTs electrode configuration.

### III. ADL Characterization

The characterization of the proposed AlScN-on-sapphire SAW ADL is carried out with a vector network analyzer (Keysight P9374A) and ground-signal-ground (GSG) probes. Short-open-load-thru calibration to the probe tips is performed to remove the influence from the measurement setup.

Fig. 6 (a) to (d) shows the measured frequency response and the extracted group delay ($\tau_G$) of the proposed ADLs with delay length ranging from 40λ to 1200λ. From the measured frequency response, the center frequency is at 5.93 GHz and the minimum IL of 7.6 dB is obtained with a delay length of 40λ, yielding a group delay 13 ns. The IL and $\tau_G$ increase linearly with the longer delay length while the passband response remains nearly unchanged across different delay lengths showing a consistent transducer performance. This also reflects the uniformity of the FBW performance with different delay length. Table I. summarizes the detailed performance parameters of the proposed AlScN-on-Sapphire ADLs showing consistent FBW performance with increased $\tau_G$. Note that all the presented frequency responses are conjugately matched with termination impedance ($Z_T$) of $26.3 + i65.5$ Ω on both ports. From the measured IL as a function of delay length, the PL can be extracted from the slope, yielding a PL performance of 9.2 dB/mm, as shown in Fig. 7(a). Similarly, by evaluating the $\tau_G$ against the delay length of each device, the group velocity ($v_g$) is extracted to be 5,778 m/s.

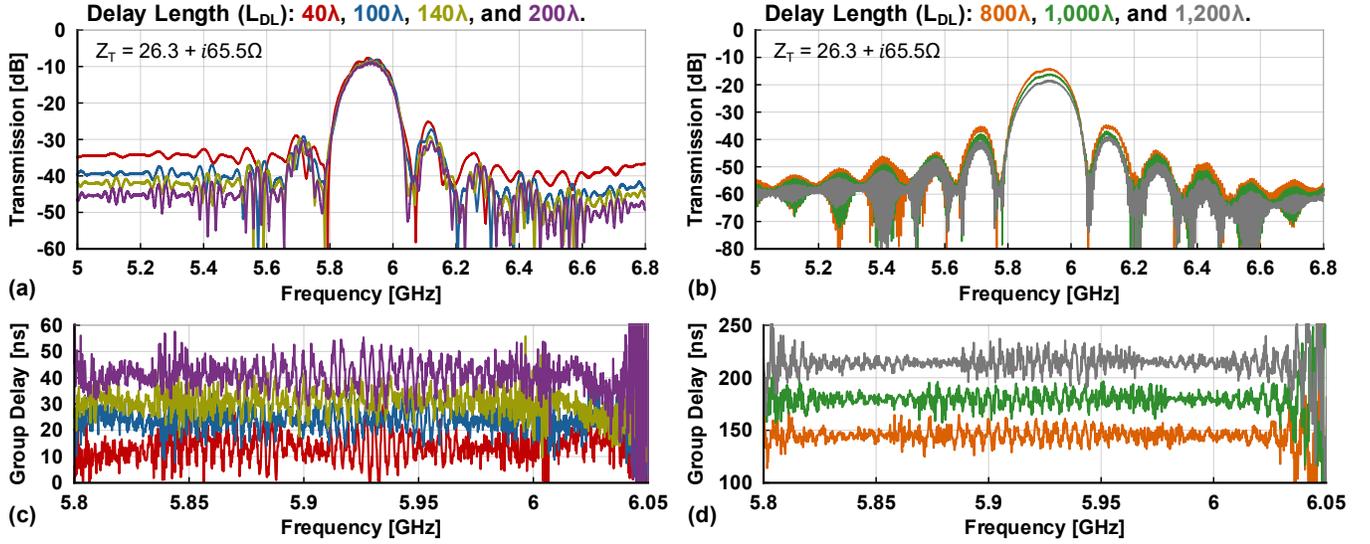

Fig. 6. (a)(b) Measured frequency response of the proposed Sezawa SAW ADL with delay length ranging from 40λ to 1200λ. (c)(d) Extracted group delay from the measured frequency response.

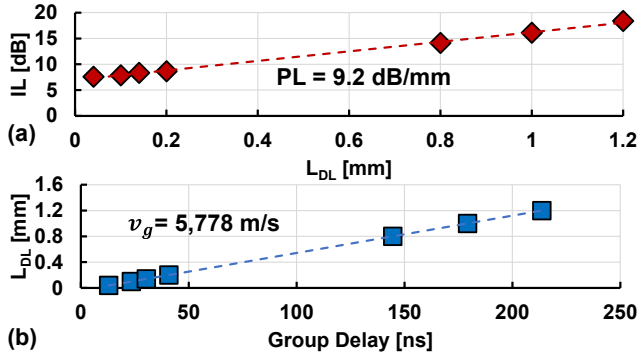

Fig. 7. (a) Extracted (a) propagation loss and (b) group velocity of the proposed ADLs.

Table. 1. Performance summary of the proposed ADLs

| $L_{DL}$ [λ] | 40 | 100 | 140 | 200 | 800 | 1,000 | 1,200 |
|---|---|---|---|---|---|---|---|
| IL [dB] | 7.6 | 7.9 | 8.3 | 8.6 | 14.1 | 16.1 | 18.4 |
| FBW [%] | 1.74 | 1.66 | 1.66 | 1.71 | 1.72 | 1.76 | 1.77 |
| $\tau_G$ [ns] | 13 | 23 | 30 | 41 | 145 | 179 | 214 |

To further assess the propagation characteristics of the AlScN-on-Sapphire Sezawa mode, the propagation $Q$-factor is calculated based on [49]. Propagation $Q$-factor consolidates the effects of wave attenuation, operational frequency, and group velocity into a single metric that reflects the intrinsic capability of the platform to support low-loss ADL. A higher propagation $Q$-factor indicates reduced propagation damping per cycle, making it a useful figure of merit for benchmarking the inherent propagation quality of the ADLs.

The propagation $Q$-factor is defined as (2)

$$Q = \frac{\pi \times f_c \times \alpha}{PL \times v_g} \quad (2)$$

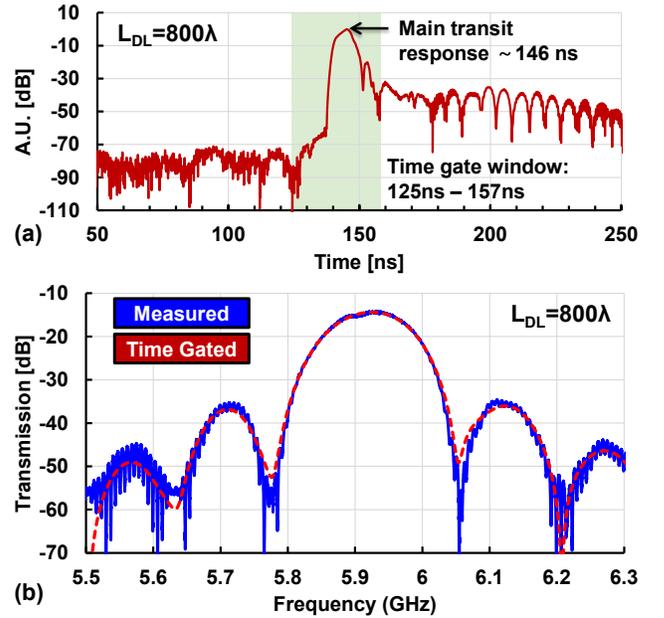

Fig. 8. (a) Time domain signal transformed by the IFFT from the frequency response. (b) Comparison chart between measured and time gated frequency response.

where $f_c$ is the center frequency in Hz, PL is the propagation loss in dB/m, and $v_g$ is the group velocity in m/s. $\alpha$ is a constant to convert the amplitude between Neper and decibel unit, as defined in (3)

$$\alpha = \frac{20}{\ln(10)} \approx 8.686 \quad (3)$$

By injecting the above-mentioned results in this work to (2), the propagation $Q$-factor of the proposed AlScN-on-Sapphire ADL could achieve up to 3,044. This high value indicates that the proposed platform has successfully confined the acoustic energy leakage which enables the low attenuation of the Sezawa modes at 5.9 GHz.

Table. 2. Comparison of SAW ADLs above 2 GHz

| Reference | $f_o$ [GHz] | $IL_{min}$ [dB] | $\tau_G$ [ns] | PL [dB/mm] | $v_g$ [m/s] | Propagation Q-factor | FBW [%] | Mode | IDT | Material |
|---|---|---|---|---|---|---|---|---|---|---|
| **This work** | **5.93** | **7.57** | **13** | **9.2** | **5,778** | **3,044** | **1.73** | **Sezawa** | **SPUDT** | **$Al_{0.6}Sc_{0.4}N$-on-Sapphire** |
| [50] | 6 | 7 | 21.8 | – | – | – | 3.14 | Shear Horizontal Composited | SPUDT | LNOI |
| [51] | 4.3 | 8.58 | 35 | – | – | – | 9.55 | Multimode Composited | SPUDT | LNOI |
| [52] | 2.7 | 5.7 | – | 6.36 | 4,320 | 2,668 | 8.4 | Shear Horizontal | SPUDT | LN-on-SiC |
| [46] | 2.5 | – | – | 4.44 | – | – | 2.59 | Sezawa | Conventional IDT | $Al_{0.7}Sc_{0.3}N$-on-Sapphire |
| [46] | 2.5 | – | – | 5.65 | – | – | 10.84 | Sezawa (etched waveguide) | Focused IDT | $Al_{0.7}Sc_{0.3}N$-on-Sapphire |
| [53] | 2.32 | 10.3 | – | 18.72 | 8,330 | §406 | N/A | Longitudinal Leaky | SPUDT | LN-on-SiC |

§ Calculated based on reported data.

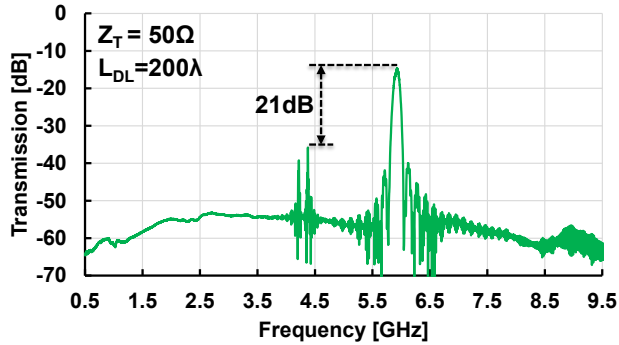

Fig. 9. Wideband frequency response covers Rayleigh and Sezawa modes under termination impedance ($Z_T$) at 50Ω without applying time gating.

To further verify the influence of the multiple reflections and the influence from the electromagnetic (EM) coupling in the proposed ADL, time gating analysis is performed on the matched frequency response. The frequency response is first transformed into time domain by an inverse fast Fourier transform (IFFT). A time gate window is then applied around the main transit response to remove the unwanted echo and EM influences. Later, the gated signal in the time domain is transformed back to the frequency domain with a fast Fourier transform (FFT). Fig. 8 shows an example of time gating based on the proposed ADL with delay length of 800λ. Fig. 8(a) shows that the main transit response after IFFT is located around 146 ns which is in agreement with the extraction from Fig. 6(d). A time gating window 125-157 ns is chosen for the FFT. The reconstructed spectrum shows almost complete overlap with the ungated response in the pass band, confirming that the triple transit echo (TTE) didn't have an obvious impact on the proposed ADLs. These results further validate that the extracted IL and propagation Q-factor represent the intrinsic propagation characteristics of the platform.

## IV. COMPARISON AND DISCUSSION

The performance of the proposed $Al_{0.6}Sc_{0.4}N$-on-sapphire Sezawa SAW ADLs is first compared with previously reported SAW ADLs above 2 GHz, as summarized in Table II. To the best of the authors' knowledge, Table II compiles the most relevant demonstrations reported in the literature for a fair comparison. At a center frequency of 5.9 GHz, the minimum IL of 7.6 dB and PL of 9.2 dB/mm are competitive with, or superior to, prior demonstrations in the range of 2-6 GHz [46], [50], [51], [52], [53]. Devices based on high-coupling piezoelectric thin films such as LN and LT can achieve significantly larger FBW while maintaining low IL, owing to their strong electromechanical coupling coefficients. Unfortunately, two previous works did not report the PL or propagation Q-factor for further comparison [50], [51]. In addition to LN and LT platforms, recent efforts in phononic integrated circuits have also explored ADLs using alternative transducer and waveguide designs [46]. For example, the etched-waveguide approach combined with focused IDTs has demonstrated large FBW and reduced PL by confining the acoustic energy in the waveguide. However, this approach comes at the cost of increased fabrication complexity and stringent process precision. In contrast, the present work achieves a propagation Q-factor exceeding 3,000 together with spectrally clean Sezawa-mode transmission, owing to the optimized thickness-to-wavelength ratio, as shown in Fig. 2(a). Fig. 9 shows an example of a wideband frequency response with $L_D$ = 200λ and $Z_T$ at 50Ω for fair comparison. The results show that the proposed Sezawa mode at 5.9 GHz has a 21 dB amplitude difference compared to the weak Rayleigh modes at 4.2-4.4 GHz.

To further benchmark the proposed ADLs, a wide range of reported ADL devices based on different technologies have been collected and compared in terms of PL, minimum IL, and propagation Q-factor versus operating frequency. The survey includes SAW-based ADLs employing shear-horizontal (SH) modes on LN and LT thin films integrated with high-velocity substrates such as SiC [52], [53], sapphire [54], and piezoelectric-on-insulator platforms [50], [51], [55], [56], [57]. Previous research on AlScN-on-sapphire Sezawa-mode ADLs have also been included in the comparison [46]. In addition to SAW ADLs, plate-wave ADLs based on suspended membranes are considered. These include antisymmetric ($A_n$) and symmetric ($S_n$) Lamb modes demonstrated on LN [4], [36], [38], [39], [41], AlScN [37], [40], [42], [58], [59], and periodic poled LN [43]. To the best of the authors' knowledge, this collection represents the most relevant demonstrations above 1 GHz and provides a fair basis for cross-platform comparison.

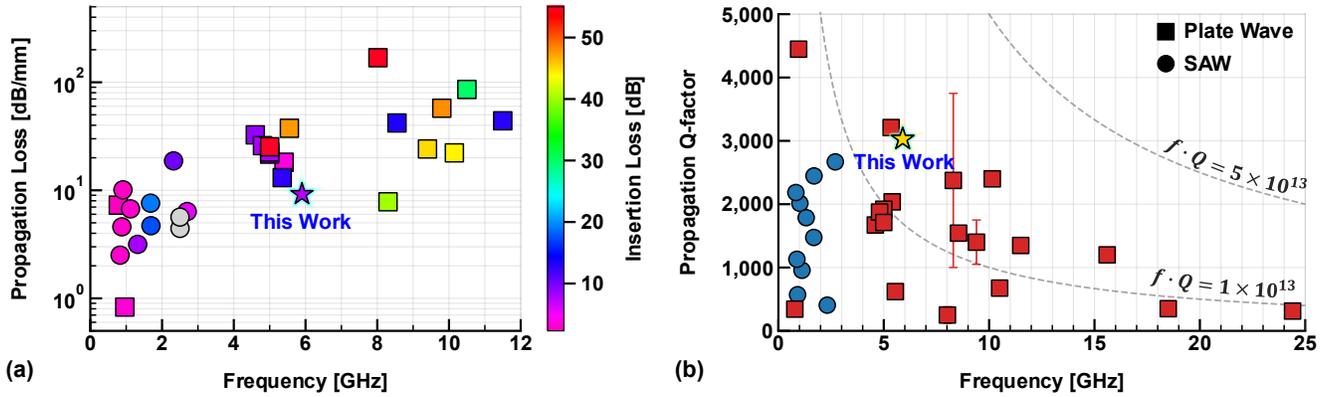

Fig. 10. (a) Comparison chart of reported PL versus operational frequency and reported minimum IL. (b) Comparison chart of reported propagation $Q$-factor versus operational frequency.

Fig. 10 summarizes the above-mentioned data by plotting the PL and propagation $Q$-factor as a function of frequency. These plots enable a direct comparison across SAW and plate-wave ADLs, highlighting the propagation trade-offs among different platforms. Fig. 10(a) shows the distribution of PL versus frequency with the reported minimum IL indicated by the color of each data point. The chart shows that PL increases with frequency for all platforms. In certain reports, plate-wave ADLs exhibit very low PL because the suspended membrane structure reduces substrate leakage. At 5.9 GHz, the proposed Sezawa-mode ADL exhibits a PL of 9.2 dB/mm, positioning itself between the reported SAW and plate-wave devices and demonstrating competitive low-loss propagation above 3 GHz. Note that some data points where only PL is available are shown in gray.

Fig. 10(b) plots the reported propagation $Q$-factor of ADLs as a function of frequency, providing a direct comparison between different platforms. The proposed $Al_{0.6}Sc_{0.4}N$-on-sapphire ADL attains a propagation $Q$-factor of 3,044 at 5.93 GHz, corresponding to an $f \cdot Q$ product of $1.8 \times 10^{13}$. To the best of the authors' knowledge, this is the first report of a propagation $Q$-factor above 3 GHz for thin-film piezoelectric SAW ADLs, marking a new benchmark for high frequency delay lines.

## V. CONCLUSIONS

In this work, we demonstrate a low-loss 5.9 GHz Sezawa mode SAW ADL on an $Al_{0.6}Sc_{0.4}N$-on-sapphire platform. By optimizing the thickness-to-wavelength ratio and employing a SPUDT electrode configuration, the devices achieved low IL of 7.6–18.3 dB across group delays ranging from 13 to 214 ns, together with an extracted PL of 9.2 dB/mm. The extracted group velocity of 5,778 m/s and propagation $Q$-factor of 3,044 corresponding to an $f \cdot Q$ product of $1.8 \times 10^{13}$, establishing a new benchmark for thin-film SAW ADLs. These results confirm that the proposed $Al_{0.6}Sc_{0.4}N$-on-sapphire configuration is a promising platform for high frequency and low-loss SAW ADLs.

## VI. ACKNOWLEDGEMENT

This work is supported by the Defense Advanced Research Project Agency under the COFFEE program. The microfabrication and microscopy were performed at the Singh Center for Nanotechnology, which is funded under the NSF National Nanotechnology Coordinated Infrastructure Program (NNCI-1542153). The authors acknowledge the use of an XRD facility supported by the Laboratory for Research on the Structure of Matter and the NSF through the University of Pennsylvania Materials Research Science and Engineering Center (MRSEC) DMR-2309043.